\documentclass[aps,prb,floatfix,preprint]{revtex4}
\usepackage{amssymb}
\usepackage{amsmath}
\usepackage{graphicx}
\usepackage{color}

\begin{document}

\title{\Huge Noisy Kondo impurities}
\author{T. Delattre$^{1,2}$, C. Feuillet-Palma$^{1,2}$, L.G. Herrmann$^{1,2}$, P. Morfin$^{1,2}$, J.-M. Berroir$^{1,2}$, G. F\`eve$^{1,2}$, B. Pla\c cais$^{1,2}$, D.C. Glattli$^{1,2,3}$, M.-S. Choi$^{4}$, C. Mora$^{1,2}$ and T. Kontos$^{1,2}$\footnote{To whom correspondence should be addressed:
kontos@lpa.ens.fr}} \affiliation{$^{1}$Ecole Normale Sup\'erieure,
Laboratoire Pierre Aigrain, 24, rue Lhomond, 75231 Paris Cedex 05,
France\\
$^{2}$CNRS UMR 8551, Laboratoire associ\'e aux universit\'es Pierre et Marie Curie et Denis Diderot, France\\
$^{3}$Service de physique de l'\'etat Condens\'e, CEA, 91192
Gif-sur-Yvette, France.\\
$^{4}$Department of Physics, Korea University, Seoul 136-713,
Korea}

\date{\today}
\maketitle

\textbf{The anti-ferromagnetic coupling of a magnetic impurity
carrying a spin with the conduction electrons spins of a host metal
is the basic mechanism responsible for the increase of the
resistance of an alloy such as Cu${}_{0.998}$Fe${}_{0.002}$ at low
temperature, as originally suggested by Kondo \cite{Kondo:64}. This
coupling has emerged as a very generic property of localized
electronic states coupled to a continuum
\cite{Crommie:98,Delley:98,Goldhaber:98,Nygard:99,Ralph:00,Liang:02}.
The possibility to design artificial controllable magnetic
impurities in nanoscopic conductors has opened a path to study this
many body phenomenon in unusual situations as compared to the
initial one and, in particular, in out of equilibrium
situations\cite{DeFranceschi:02,Marcus:05,Grobis:08}. So far,
measurements have focused on the average current. Here, we report on
\textit{current fluctuations} (noise) measurements in artificial
Kondo impurities made in carbon nanotube devices. We find a striking
enhancement of the current noise within the Kondo resonance, in
contradiction with simple non-interacting theories. Our findings
provide a test bench for one of the most important many-body
theories of condensed matter in out of equilibrium situations and
shed light on the noise properties of highly conductive molecular
devices.}

The hallmark of the Kondo effect in a quantum dot is an increase
of the conductance below $T_K$ up to the unitary conductance
$2e^{2}/h$ at very low temperature and bias voltage. This
corresponds to the opening of a spin degenerate conducting channel
of transmission $1$ at the Fermi energy of the electrodes, if only
a single spin 1/2 is involved. The non-interacting theory of shot
noise predicts no noise for such a quantum scatterer, as a
consequence of Fermi statistics \cite{Blanter:00}. Does such a
statement apply to a generic Kondo resonance ?

In this letter,  we show that, in contrast to the prediction of the
non-interacting theory, a conductor in the Kondo regime can be noisy
even though its conductance is very close to $2e^2/h$. We report on
noise measurements carried out in single wall carbon nanotube based
quantum dots in the Kondo regime. We find an enhancement up to an
order of magnitude with respect to the non-interacting (Fermi gas)
theory of the shot noise within the Kondo resonance due to charge
quantization combined with spin and orbital degeneracy. We can
account for this enhancement with a fully interacting theory based
on the Slave Boson Mean Field technique (SBMF). We also find a
non-monotonic variation of the equilibrium current fluctuations as
the temperature is decreased below $T_K$. Finally, the conductance
\textit{and} the noise obey a scaling law, the Kondo temperature
$k_B T_K$ being the only energy scale.

Single wall carbon nanotubes (SWNTs) are ideally suited to explore
the noise in the Kondo regime of a quantum dot. They can exhibit
Kondo temperatures up to $10-15K$, which allows to apply rather high
currents, up to $10 nA$, \textit{within} the Kondo resonance. This
contrasts with semiconducting quantum dots where Kondo temperatures
are usually much smaller and therefore shot noise measurements more
difficult (a specific aspect of the spin 1/2 case could be studied
only very recently in a lateral quantum dot \cite{Zarchin:08}). In
addition, carbon nanotubes allow to investigate a large class of
different Kondo effects, including the simplest spin 1/2 Kondo
effect \cite{Nygard:99}, the 2-particle Kondo effect \cite{KK2}, the
orbital Kondo effect \cite{SU4exp1} and the so-called SU(4) Kondo
effect\cite{SU4exp1,SU4exp3}.

We first present results for device A which consists of a SWNT
contacted by two Pd electrodes separated by $200nm$ (see figure
\ref{greyscale}b). Our measurement setup, shown on figure
\ref{greyscale}b, allows to measure simultaneously the noise and the
conductance. The \textit{total current noise} $S_I$ of the SWNT is
obtained by substracting a calibrated offset to the measured
cross-correlations (see Methods and Supplementary information). The
color scale plot of the differential conductance of the sample is
displayed on figure \ref{greyscale}a and the characteristic Kondo
ridges are observed at $1.4K$ as the horizontal lines within the
Coulomb diamonds. We now study the two
 ridges, Sample A Ridge 1 (SAR1) and Sample A Ridge 2
(SAR2), corresponding to the gate voltage $V_G = 11.26V$ and $V_G =
11.80V$ respectively, indicated by an arrow in figure
\ref{greyscale}a.

Signatures of the Kondo effect are already present in the
equilibrium current fluctuations. From the fluctuation-dissipation
theorem, the power spectral density of current noise is expected to
be given by the Johnson-Nyquist formula : $S_I(V_{sd}=0)=4 k_B T
G(T,V_{sd}=0)$, where $G(T,V_{sd})=dI/dV$ is the differential
conductance at temperature $T$ and source-drain bias $V_{sd}$.
However, due to Kondo correlations, $G(T,V_{sd}=0)$ displays a sharp
increase as $T$ is lowered, as shown on figure \ref{greyscale}c left
inset, in blue dashed lines. From the Johnson-Nyquist formula, an
increase of $G(T,V_{sd}=0)$ tends to produce an increase of $S_I$ as
the temperature is lowered but at zero temperature, $S_I$ should be
zero. Therefore, an unusual maximum can occur in the variation of
$S_I$ as a function of temperature for a quantum dot in the Kondo
regime. We observe such a maximum around $3K$ for ridge SAR1, as
shown in the left inset of figure \ref{greyscale}c in black squares,
which reaches about $7$ to $9 \times 10^{-27} A^{2}/Hz$ here. As
expected, the Johnson-Nyquist relationship still holds, as shown in
the right inset of figure \ref{greyscale}c.

The bias dependence of $G(1.4K,V_{sd})$ for SAR1 and SAR2 is
displayed on figure \ref{noise}. The characteristic zero bias peak
of the Kondo effect is observed. The conductance at zero bias is,
for both cases, close to $2e^2/h$ (respectively $0.997 \times
2e^2/h$ and $0.865 \times 2e^2/h$). The half-width of these
resonances, of about $0.2meV$, gives an estimate of the Kondo
temperature, of about $2.5K$, consistent with the temperature
dependence shown in figure \ref{greyscale}c left inset. As shown on
figure \ref{noise}a and b (black squares), when a finite bias is
applied, the \textit{total} noise power spectral density $S_I$
increases from about $5 \times 10^{-27} A^{2}/Hz$ to about $8 \times
10^{-27} A^{2}/Hz$ ($7 \times 10^{-27} A^{2}/Hz$) for SAR1(SAR2)
respectively. The ratio of $S_I$ to the Schottky value $2eI_{sd}$,
where $I_{sd}$ is the current flowing through the nanotube at
$V_{sd}=0.84mV$, is about $0.84 \pm 0.09$($0.89 \pm 0.1$) for
SAR1(SAR2) respectively. Therefore, the noise remains
sub-poissonian.

In general, the noise properties of carbon nanotubes are affected by
the existence of a possible orbital degeneracy, which arises from
the band structure of graphene, as recently shown in the
non-interacting limit \cite{Delattre:07}. Therefore, a first step
towards the understanding of the measurements presented in figure
\ref{noise} is to use a resonant tunneling model with \textit{two
spin degenerate channels}, with transmission
$\widetilde{D}_{i,res}(\epsilon)=d_i /(1+\epsilon^2/\Gamma^2)$,
$d_i$ being the transmissions of the channel of index $i\in\{1,2\}$
and $\Gamma$ being the width of the resonant level and $\epsilon$
being the energy. From the non-interacting scattering theory
\cite{Blanter:00}, the current and the noise associated with
$\widetilde{D}_{i,res}$ read:
\begin{eqnarray}
I(V_{sd})=\frac{2e}{h}\sum_{i=1,2}\int_{-\infty}^{\infty}\widetilde{D}_{i,res}(\epsilon)(f_L-f_R)d\epsilon \label{eq:currentSBMF}\\
S_{I}(V_{sd})= \frac{4e^2}{h} \sum_{i=1,2}\int d\epsilon \big \{
 \widetilde{D}_{i,res}(\epsilon)[f_L(1-f_L)+f_R(1-f_R)] \label{eq:noiseSBMF}\\
+\widetilde{D}_{i,res}(\epsilon)[1-\widetilde{D}_{i,res}(\epsilon)](f_L-f_R)^2\big\}\nonumber
\end{eqnarray}
with $f_L=f(eV_{sd}/2+\epsilon)$ and $f_R=f(-eV_{sd}/2+\epsilon)$,
$f(\epsilon)$ being the Fermi function at temperature $T$. The
fits of $dI/dV$ using equation (\ref{eq:currentSBMF}) and
$\widetilde{D}_{i,res}$, shown in blue dashed lines in figure
\ref{noise}, panel a.(b.), yield $d_1 = d_2 = 0.95$ ($d_1 = d_2 =
0.99$) and $\Gamma=0.11 meV$ ($\Gamma=0.09 meV$) respectively.
These fits are poor because the Lorentzian line shape with
constant $\Gamma$ assumed for $\widetilde{D}_{i,res}$ is not able
to account for both the height and the width of the measured
dependence of $dI/dV$ as a function of $V_{sd}$. Furthermore, the
noise obtained with formula (\ref{eq:noiseSBMF}) using the above
values for $d_1$, $d_2$ and $\Gamma$, in blue dashed lines in the
lower panels of figure \ref{noise}, is about an order of magnitude
smaller than our experimental findings. Therefore, a simple
non-interacting resonant tunneling theory can account neither for
the conductance nor for the noise that we observe.

In the Kondo regime, in case of a fourfold degeneracy and single
charge occupancy, the maximum of the resonance lies at $T_K$ above
the Fermi energy of the leads according to the Friedel sum rule, as
depicted on figure \ref{cartoon}b. From this, one can infer that, if
the couplings of the level to the left $\Gamma_L$ and the right
$\Gamma_R$ electrodes are the same (hereafter called the symmetric
case), the differential conductance which saturates at $2e^2/h$
corresponds to \textit{two channels of transmission 1/2}
($\frac{1}{2}\times\frac{4\Gamma_L\Gamma_R}{(\Gamma_L+\Gamma_R)^2} $
in the general case). This corresponds to the so-called SU(4) Kondo
effect where the spin \textit{and} the orbital degree of freedom
play an equivalent role \cite{MahnSoo:06,Simon:07} in the Kondo
screening.

Unfortunately, no full out-of-equilibrium theory of the Kondo effect
is available. As shown below, our experiments are carried out in a
regime where $T\sim T_K/3$ and $eV_{sd} \lesssim 3 k_BT_K$.
Therefore, one has to choose a low energy theory in order to
understand our experiment. Essentially 3 different approaches can be
used : the Fermi Liquid (FL) theory, the Slave Boson Mean Field
(SBMF) theory and the Non-Crossing Approximation
(NCA)\cite{Golub:02}. Both SBMF and FL theory are expected to be
exact in the zero-energy limit ($k_BT=eV_{sd}=0$). Presently
available FL theories \cite{Komnik:04,Sela:05,Mora:08,LeHur:08}
provide the correct description only at very low energies and give
unphysical results for the temperature and bias range of our
experiment.  On the other hand, the SBMF theory turns out be more
robust up to the energy range of our experiment \cite{Golub:02}. The
NCA is a good approximation at relatively higher energies, and
becomes inaccurate at temperatures much smaller than $T_K$.  It
gives results similar to the SBMFT at energies not too small
compared with $T_K$ \cite{Golub:02}. Therefore, we use a SBMF
approach which is the simplest one available. In order to account
for the orbital degeneracy, we use a SU(4) theory. Such a model
should be regarded as the minimal one to explain our data and one
should bear in mind that our samples might be in a regime where the
full fourfold degeneracy is only approximately fulfilled. The SBMF
has been widely used in the spin degenerate (SU(2)) case to compute
the noise in quantum dots in the Kondo regime\cite{Lopez:04} and in
the SU(4) case to study the conductance \cite{MahnSoo:06}. We
generalize here this approach for the noise. In the SBMF approach,
one still uses formulae (\ref{eq:currentSBMF}) and
(\ref{eq:noiseSBMF}) replacing $\widetilde{D}_{i,res}(\epsilon)$ by
$\widetilde{D}_{SBMF}(\epsilon,V_{sd})$ which accounts  for the
interactions in a self-consistent way. It is a Breit-Wigner formula
with a level position $\widetilde{\epsilon_0}$ and a width
$\widetilde{\Gamma}$ which explicitly depend on $V_{sd}$ and $T$. In
the SU(4) limit, one has in the symmetric case:
\begin{eqnarray}\label{eq:res}
\widetilde{D}_{SBMF}(\epsilon,V_{sd},T)=
\frac{\widetilde{\Gamma}^2}{(\frac{\epsilon}{k_B T_K}
-\widetilde{\epsilon_0})^2+\widetilde{\Gamma}^2}
\end{eqnarray}
with $\widetilde{\Gamma}^2 \approx
-\frac{t^2}{6}-\frac{x^2}{8}+\sqrt{1+(\frac{t^2}{6}+\frac{x^2}{8})^2}$
and
${\widetilde{\epsilon_0}}^2\approx\frac{t^2}{6}+\frac{x^2}{8}+\sqrt{1+(\frac{t^2}{6}+\frac{x^2}{8})^2}$,
where $x=eV_{sd}/k_B T_K$, $t=\pi T/T_K$.

From formula (\ref{eq:res}),
$\widetilde{D}_{SBMF}(\epsilon=0,V_{sd}=0,T=0)=1/2$. This contrasts
with the general expression of $\widetilde{D}_{i,res}(\epsilon)$ for
which the transmission at zero energy can take any value between $0$
and $1$. In addition, $\widetilde{\epsilon_0}$ and
$\widetilde{\Gamma}$ are universal functions of $x=eV_{sd}/k_B T_K$
and $t=\pi T/T_K$. Both facts originate from electron-electron
interactions. From equations (\ref{eq:currentSBMF}) and
(\ref{eq:res}), the conductance is fitted with \textit{only one}
parameter, $k_B T_K$. The outcome of the combination of equations
(\ref{eq:currentSBMF}), (\ref{eq:noiseSBMF}) and (\ref{eq:res}) is
shown as solid red lines in figure \ref{noise}a and figure
\ref{noise}b for $k_B T_K=0.305 meV$ and $k_B T_K=0.26meV$
respectively.  We find a good agreement with the experimental data
using this SU(4) theory. The agreement is quantitative for the noise
of SAR1 and the conductance of SAR2. The conductance of SAR1 peaks
at a higher value than the one predicted by the SBMF, although the
theoretical line corresponds to the fully symmetric case. This is
probably due to the fact that the sample is not far from the mixed
valence regime \cite{Haldane:78} for SAR1. Indeed, as can be seen on
the linear conductance curve of figure \ref{greyscale}a., the single
charge peaks slightly overlap at $V_G = 11.26V$. The noise of SAR2
is close to the theoretical curve in the (most important) out of
equilibrium regime i.e. $\mid V_{sd}\mid
>0.2meV$. For low bias, a spurious shift less than
$10\%$ of the total noise, of about $0.4 \times 10^{-27} A^{2}/Hz$,
occurs. It probably originates from a systematic background
variation for $V_G=11.80V$. Overall, the SBMF SU(4) theory is in
much better agreement than the non-interacting resonant tunneling
theory. Both the conductance \textit{and} the noise data are
accounted for by a \textit{single} energy scale, $k_B T_K$, even
though the two actual Kondo impurities have Kondo temperatures
differing by about $15\%$.

A scaling behavior is a well-known property of the Kondo
effect\cite{Haldane:78}. It has been tested only very recently in a
lateral quantum dot, for the bias dependence of the conductance in
the SU(2) case \cite{Grobis:08}. We have studied 4 ridges in device
A and 1 in device B. For the 5 different Kondo ridges studied, we
observe a scaling of $G(T,0)$ versus $T/T_K$ and of $G(1.4,V_{sd})$
versus $eV_{sd}/k_B T_K$, as shown on figure \ref{universal}a and b.
For the temperature and bias dependences, the data is very well
accounted for by the empirical formulae
$G=C_{0}/(1+(2^{1/s}-1)(T/T_K)^2)^{s}$ \cite{Grobis:08,SU4exp2} with
$s=1.02 \pm 0.04$ and $G=G(1.4K,0)\times \exp[-2(\frac{eV_{sd}}{2.3
k_BT_K})^2]$ respectively. Independently of the low energy theory, a
similar behavior is expected for the noise
\cite{Golub:02,Komnik:04,Sela:05,LeHur:08,Mora:08} because scaling
is an essential feature of Kondo physics, linked to the fact that
there is only one energy scale, $T_K$ which controls the electronic
system. We introduce $I_0=\frac{2e^2}{h} \times 2 D_0 V_{sd}$ and
$S_0=\frac{4e^2}{h}(4 k_B T {D_0}^2+2e
D_0(1-D_0)V_{sd}\coth(\frac{eV_{sd}}{2k_B T}))$ which are the zeroth
order for the current and the noise for $T/T_K, eV_{sd}/k_B T_K
\rightarrow 0$ in the theory. For each Kondo ridge, the parameters
$D_0$ ($D_0=\frac{h}{4e^2} G(T=0,V_{sd}=0)=2\frac{\Gamma_L
\Gamma_R}{(\Gamma_L +\Gamma_R)^2}$) and $T_K$ are obtained from the
fitting of the conductance with the SBMF theory. We get the
corresponding sets $\{\text{Sample}, \text{Gate voltage}, k_B
T_K/e,2D_0\}$ as $\{A, 11.80V, 0.26mV,1\}$, $\{A, 9.67V,
0.181mV,0.94\}$, $\{A, 17.65V, 0.26mV,0.58\}$, $\{A, 11.26V,
0.305mV,1\}$ and $\{B, 3.025V, 0.29mV,0.99\}$. In order to test the
scaling of the noise, we study $\delta S=S_{exc,0}-S_{exc,I}$ as a
function of $\delta I=I_0-I$ as displayed on figure
\ref{universal}c, $S_{exc,(0,I)}$ being the excess noise defined as
$S_{exc,(0,I)}=S_{(0,I)}(V_{sd})-S_{(0,I)}(V_{sd}=0)$. We observe
the same behaviour for all the 5 ridges which can be fitted as
 $\delta S = (0.45 \pm 0.05) \times 2e\delta I + 0.2 \pm 0.2$. The value of
 the slope is the central \textit{quantitative} result of this letter.
It is close to $1/2$ which is the number predicted by the SBMF
theory.  For the symmetric case, this value is exact at $T=0$ and
 is a very good approximation
 up to $T=T_K/2$ (see Supplementary information). Even though $T_K$
and/or $D_0$ can vary by as much as $40\%$, the noise properties of
the Kondo impurity remain invariant.

\textbf{METHODS}\\
\textbf{Experimental} \\
Our SWNTs are grown by chemical vapor deposition. They are localized
with respect to alignment markers with an atomic force microscope
(AFM) or a scanning electron microscope(SEM). The contacts are made
by e-beam lithography followed by evaporation of a $70nm$-thick Pd
layer at a pressure of $10^{-8} mbar$. The highly doped Si substrate
(resistivity of $4-8 m \Omega.cm$) covered with $500nm$ $SiO_{2}$ is
used as a back-gate at low temperatures. The typical spacing between
the Pd electrodes is between $200nm$ and $500nm$. The current
fluctuations in the NT result in voltage fluctuations along the two
resistors (of $200\Omega$) shown in figure \ref{greyscale}b. The two
signals are fed into two coaxial lines and separately amplified at
room temperature by two independent sets of low noise amplification
stages (gains: $G_{1}=7268 \pm 10$, $G_{2}=7304 \pm 10$, amplifiers
NF SA-220F5). We calculate the cross correlation spectrum with a
spectrum analyzer. Each noise point corresponds to 20 averaging runs
of 40000 spectra with a frequency span of $78.125 kHz$ (1601
frequency points) and a center frequency of $1.120MHz$ or
$2.120MHz$. The sensitivity is about $3\times 10^{-28} A^2/Hz$. The
raw data are corrected by an offset amounting to $(4.2 (\pm 0.1) +
19.5 (\pm 0.2) \frac{h}{4e^2} G(T,V_{sd}))10^{-27} A^2/Hz$ which is
in very good agreement with the circuit diagram of our
measurement setup (see Supplementary information). \\
\textbf{Fitting details}\\
 Throughout the paper, a SU(4) SBMF theory is
used (as a minimal model). However, the SU(2) SBMF theory does not
account for the data (see supplementary information). When fitting
with the SBMF, we have interpolated the low bias expression of
$\widetilde{\Gamma}$ with a $V_{sd}$-independent expression for
$|V_{sd}|>0.7meV$, $\widetilde{\epsilon_0}$ remaining unchanged.
Such an interpolation has been widely used (see e.g.
\cite{Aguado:04}) in order to cope with the well-known phase
transition problem of the SBMF approach at high bias (see
Supplementary information). Finally, when fitting the temperature
dependence of the conductance for all the 5 ridges studied with the
empirical formula described in the main section, we have found a $s$
parameter larger than that usually found in the litterature except
for the experiment in reference 17, where the SU(4) case was
considered for the first time. Note however that it is distinct from
the non-interacting resonant tunneling model which would lead to
$1/T$ dependence for large $T$.

{\noindent\small{\bf Correspondence. } Correspondence and requests
for materials should be addressed to T.K. : kontos@lpa.ens.fr

{\noindent\small{\bf Acknowledgements. } We thank A. Cottet for a
critical reading of the manuscript and K. LeHur, P. Simon, L.
Glazman and N. Regnault for illuminating discussions. This work is
supported by the SRC (R11-2000-071) contract, the BK21 contract, the
ANR-05-NANO-055 contract, the EU contract FP6-IST-021285-2 and by
the C'Nano Ile de France contract SPINMOL.

{\noindent\small{\bf Competing financial interests. } The authors
declare no competing financial interests.

\begin{figure}[!htph]
\caption{\textbf{Anomalous temperature dependence of the current
noise at the onset of the Kondo effect.}\\ a. Colorscale plot of the
differential conductance as a function of the gate voltage $V_G$ and
the source-drain bias $V_{sd}$ at $T=1.4K$. The characteristic
horizontal lines in the middle of the Coulomb diamonds signaling the
Kondo effect are observed. In green lines, the linear conductance
curves at $T=1.4K$ (solid line) and $T=12K$ (dashed line). b.
Simplified diagram of the noise measurement scheme and SEM picture
of a typical sample. The bar corresponds to $1\mu m$. c. In the left
inset, the non-monotonic temperature dependence of the equilibrium
current fluctuations on the Kondo ridge SAR1 ($V_{G}=11.26V$ (black
squares)). In blue circles, the corresponding variation of
$G(T,V_{sd}=0)$. In solid
 black line, the predicted dependence of $S_I$ from the Johnson-Nyquist
formula. In the right inset, $S_I$ is plotted versus $4k_B T
G(T,V_{sd}=0)$. The line corresponds to the expected slope of $1$.
The error bars correspond to the mean square root of the statistical
error and the systematic error due to fluctuations of
the background.}%
\label{greyscale}%
\end{figure}

\begin{figure}[!htph]
\caption{\textbf{Noise enhancement within the Kondo resonance.}\\
a. Conductance and noise measurements (black circles and black
squares respectively) for the Kondo ridge 1 of Sample A
($V_{G}=11.26V$) at $1.4K$.  The results of the SBMF theory, in
solid red lines, are in good agreement with the conductance and the
noise. b. Similar plots as in panel a. for the Kondo ridge 2 of
sample A ( $V_{G}=11.80V$). The Kondo resonance corresponds to a
slightly different Kondo temperature. In both panels, the
non-interacting theory, in blue dashed lines, accounts neither for
the conductance nor for the noise. The error bars correspond to the
mean square root of the statistical error and the systematic error
due to fluctuations of
the background.}%
\label{noise}%
\end{figure}

\begin{figure}[!htph]
\caption{\textbf{Schematics of the two limiting cases for noise in a
single wall carbon nanotube in the Kondo regime.}\\ The blue line
depicts the line shape of the energy dependent transmission of the
Kondo resonance. The red line depicts the part of the transmission
accessible in the "transport window" $eV_{sd}$. The rose dashed line
is the position of the Fermi level of the reservoirs for $V_{sd}=0$.
a. For the twofold degenerate case, the effective transmission per
channel is close to $1$ and only one channel contributes to
transport, leading to a suppression of the shot noise ($D(1-D)$
close to $0$). b. For the fourfold degenerate case, the effective
transmission per channel is close to $1/2$ and there are two
channels, leading to an enhancement of the shot noise ($D(1-D)$
close to $1/4$).
}%
\label{cartoon}%
\end{figure}

\begin{figure}[!htph]
\caption{\textbf{Scaling properties of the noise of the Kondo 'impurities'.}\\
a. Scaling of the temperature dependence of the zero bias
conductance. b. Scaling of the bias dependence of the conductance at
$1.4K$. For the temperature (panel a.) and bias dependence (panel
b.), the data is very well accounted for by the empirical formulae
$G=C_{0}/(1+(2^{1/s}-1)(T/T_K)^2)^s$ with $s=1.02$ and
$G=G(1.4K,0)\times \exp[-2(\frac{eV_{sd}}{2.3 k_B T_K})^2]$
respectively (in solid lines). c. Scaling of the noise for the
different 'Kondo impurities' measured. The line corresponds to the
slope of $1/2$ predicted by the SBMF. Inset: the different symbols
used in the figure and their corresponding 'Kondo impurity'.
}%
\label{universal}%
\end{figure}

\clearpage 
\centering\includegraphics[height=0.55\linewidth,angle=0]{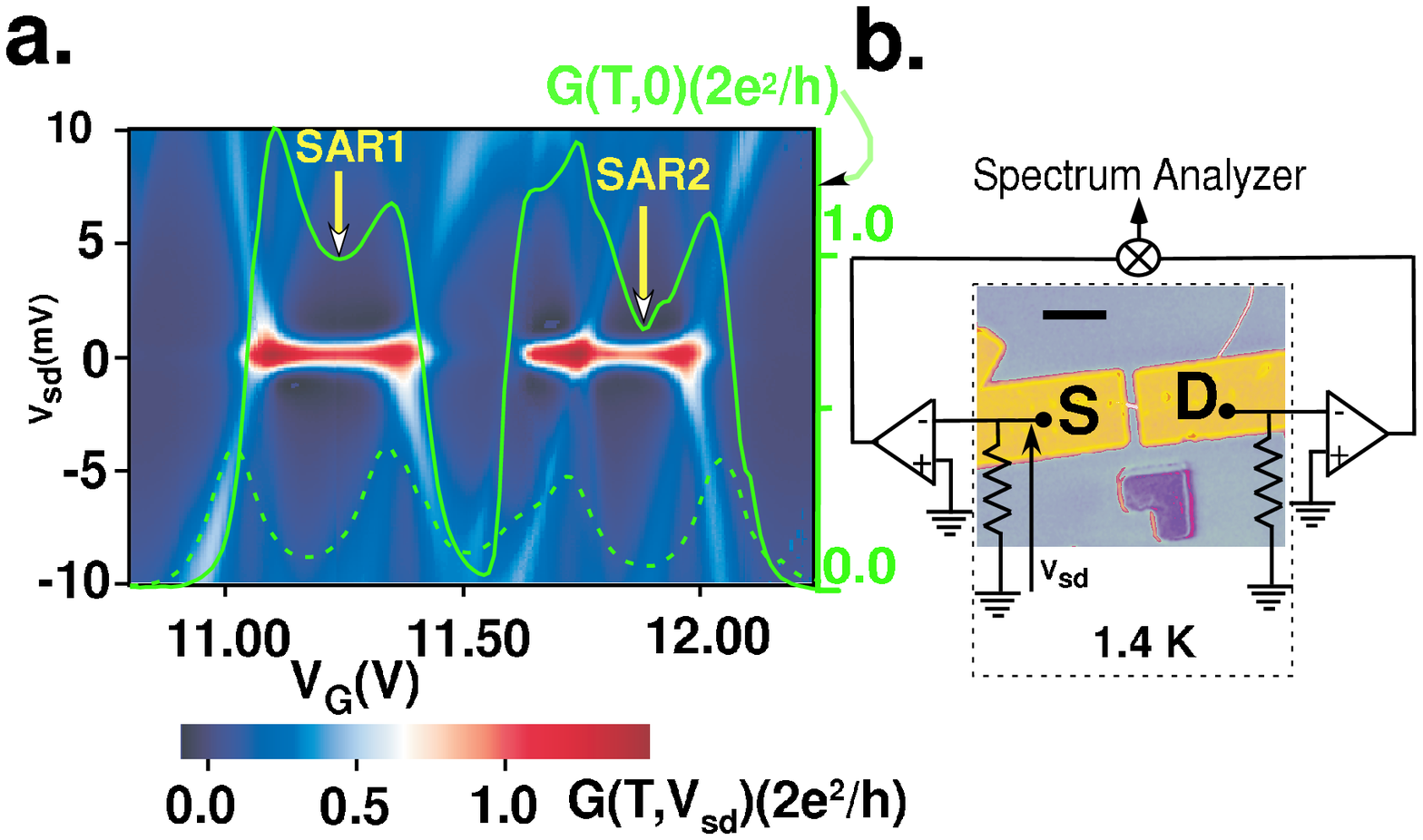}\quad\includegraphics[height=0.65\linewidth,angle=0]{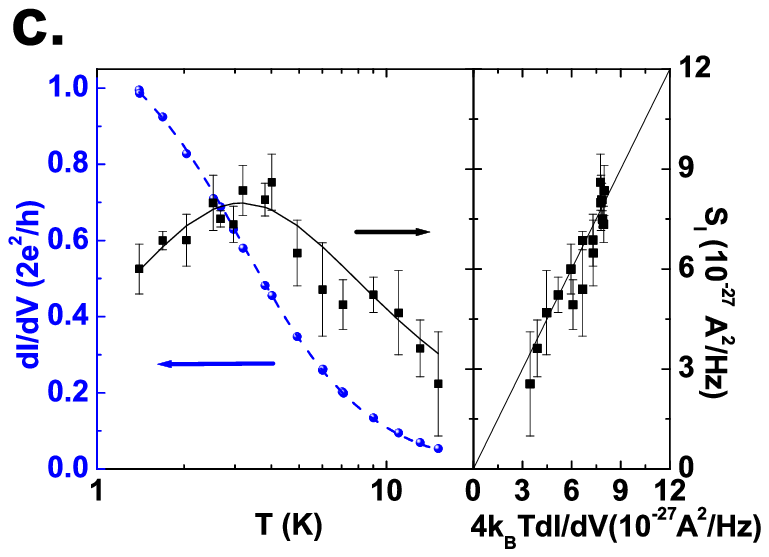}
\newpage
\centering\includegraphics[height=1.2\linewidth,angle=0]{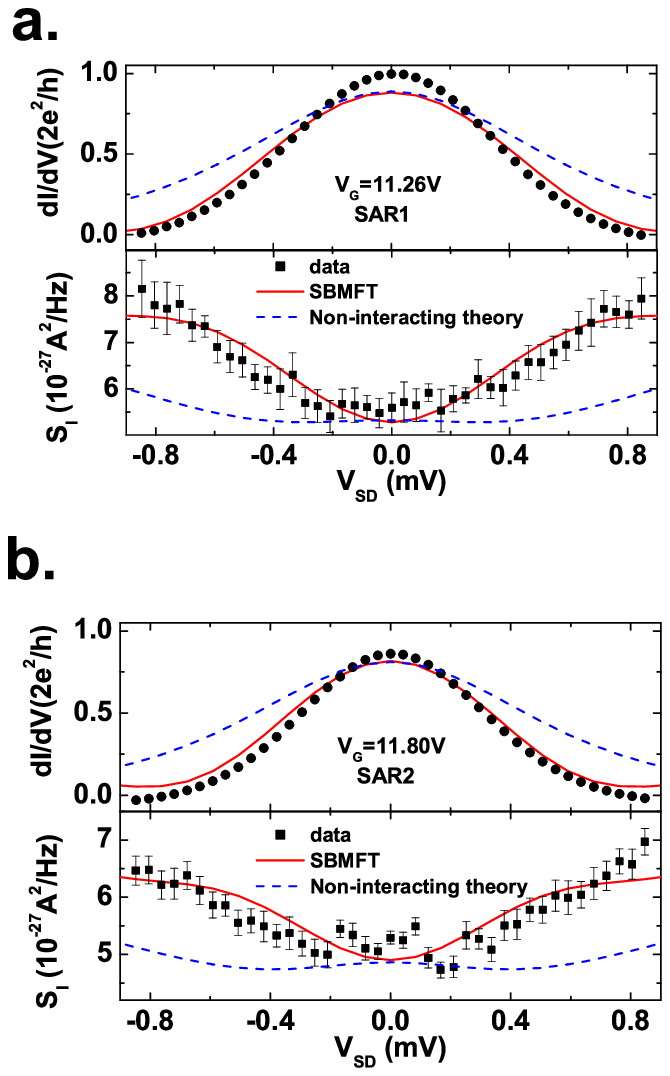}
\newpage
\centering\includegraphics[height=0.5\linewidth,angle=0]{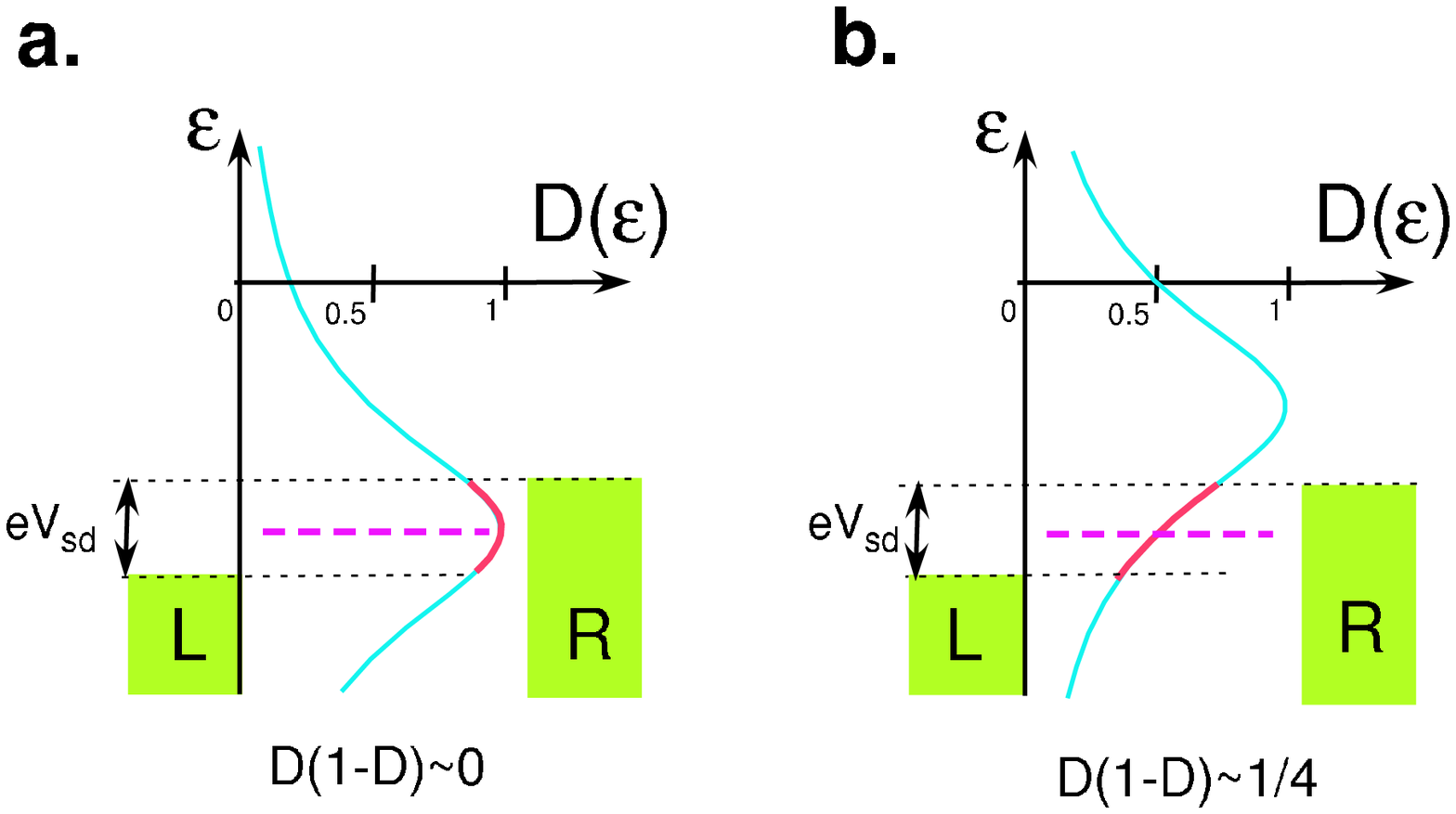}
\newpage
\centering\includegraphics[height=0.99\linewidth,angle=0]{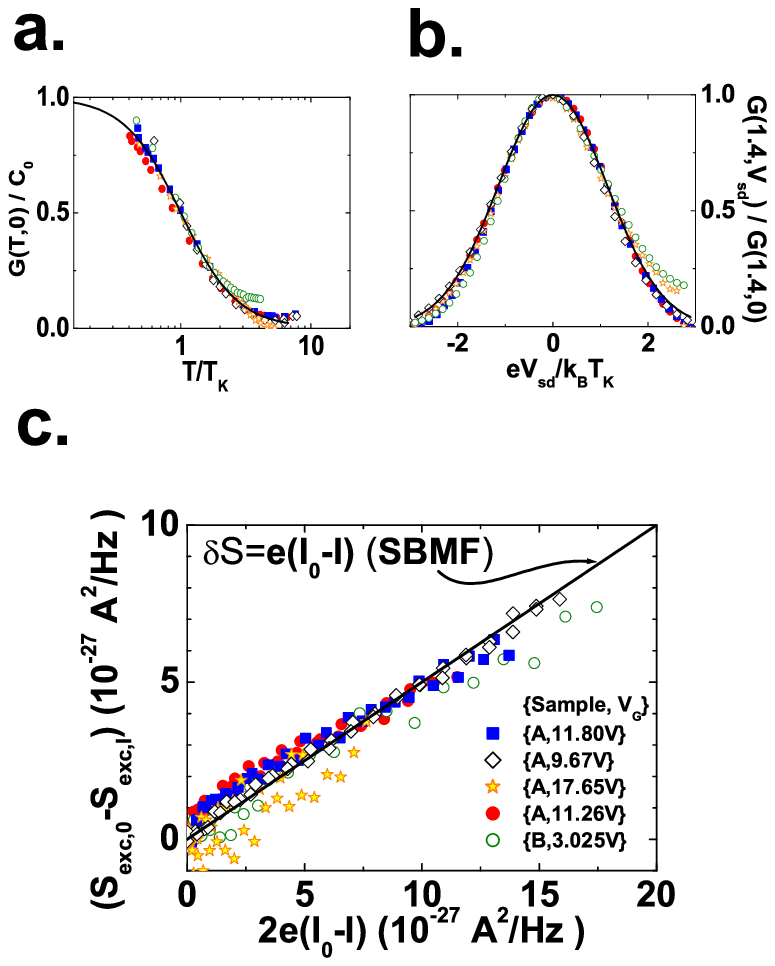}

\end{document}